\documentclass{article}

\usepackage[dblblindworkshop, final]{neurips_2026}
\workshoptitle{GlobalSouthAI}
\usepackage[utf8]{inputenc}
\usepackage[T1]{fontenc}
\usepackage{hyperref}
\usepackage{url}
\usepackage{booktabs}
\usepackage{amsmath,amssymb}
\usepackage{microtype}
\usepackage{graphicx}
\usepackage{xcolor}
\usepackage{tikz}
\usepackage{siunitx}
\usetikzlibrary{arrows.meta,positioning,fit,calc}

\definecolor{tg}{RGB}{102,145,187}
\definecolor{attn}{RGB}{215,128,171}
\definecolor{ctrl}{RGB}{235,163,73}
\definecolor{lm}{RGB}{61,166,123}
\definecolor{mem}{RGB}{142,119,181}

\title{\textsc{PENSA}: A Spatial Accelerator for BitNet~b1.58 Inference on an HBM-equipped AMD Alveo FPGA}
\author{Karthikeyan B. Kalyanasundaram$^1$ \quad Ivan S.~Maksymov$^1$ \quad Philip H.W. Leong$^{1,2}$\\
$^1$
The University of Sydney \quad
$^2$TernaryNet\\
\texttt{\{karthikeyan.kalyanasundaram, ivan.maksymov,
philip.leong\}@sydney.edu.au}}

\begin{document}
\maketitle

\begin{abstract}
Ternary large language models (tLLMs) replace general multiplication in dominant linear layers with sign-controlled accumulation, shifting hardware bottlenecks toward packed-weight delivery, reduction, and operator occupancy. \textsc{PENSA} is an end-to-end FPGA accelerator for the BitNet~b1.58~2B4T model on an AMD Alveo U50\texttrademark. Two row-partitioned ternary matrix engines consume packed TQ1\_0 weights from eight independent high-bandwidth memory (HBM) pseudo-channels, each processing 1,280 ternary weights per cycle through sign selection and integer reduction. The sequencer, attention, and vocabulary-projection units remain resident on the FPGA, while two-stream interleaved scheduling overlaps sequences across the persistent pipeline to form the Penta-Unit. At 104.1~MHz, \textsc{PENSA} achieves 58.09~tok/s at context 128 and 24.60~tok/s at context 2048, corresponding to 2.14--3.14$\times$ the throughput of a same-model Xeon E-2356G baseline. The resulting system demonstrates end-to-end FPGA execution of 2B-scale BitNet topology without modifying its model structure.
\end{abstract}

\section{Introduction}

Ternary large language models (tLLMs) constrain linear-layer weights to \(w\in\{-1,0,+1\}\), replacing general matrix multiplication with conditional addition/subtraction while reducing weight-storage and memory-bandwidth requirements~\cite{ma2024era,ma2025bitnet}. Software implementations such as Bitnet.cpp specialize ternary inference through lookup-table (LUT) and Multiply/Add-based CPU kernels~\cite{bitnetcpp}. However, removing general multiplication does not make end-to-end inference uniformly compute-light: the limiting work shifts toward packed-weight delivery, reduction, operator occupancy, attention, and vocabulary projection.

FPGA accelerators explore complementary forms of ternary specialization. TeLLMe and TeLLMe~v2 employ lookup-based ternary matrix multiplication~\cite{qiao2025tellme,qiao2026tellmev2}. TENET combines sparse LUT computation with dynamic activation N:M sparsity, weight decompression, and sparse-attention dataflow~\cite{huang2025tenet}, while TerEffic explores ternary compute--memory co--design~\cite{yin2025tereffic}. Other designs employ sparse or LUT-oriented ternary datapaths~\cite{zhu2026sparsegemm,geens2026lut,pittet2026ssr}, while FPGA LLM accelerators more broadly exploit spatial pipelines and specialized memory organizations~\cite{chen2023spatial,zheng2025looplynx,zhuang2024ssr,zeng2024flightllm}. However, end-to-end FPGA inference of BitNet~b1.58~2B4T with direct packed-weight execution, high-bandwidth memory (HBM)-fed ternary engines, persistent heterogeneous operators, and inter-sequence overlap remains comparatively unexplored.

\textsc{PENSA} targets this integration point as an end-to-end platform for characterizing the consequences of native ternary execution for BitNet~b1.58~2B4T. It consumes packed ternary weights from HBM, partitions linear projections across two persistent TGEMMs, and keeps autoregressive execution resident on the FPGA. The resulting measurements expose how bottlenecks shift from multiplication toward data delivery and reduction at the ternary engines, persistent-unit occupancy at short contexts, and sequence-dependent attention as context length increases.
\begin{figure}[ht]
\centering
\includegraphics[width=\linewidth]{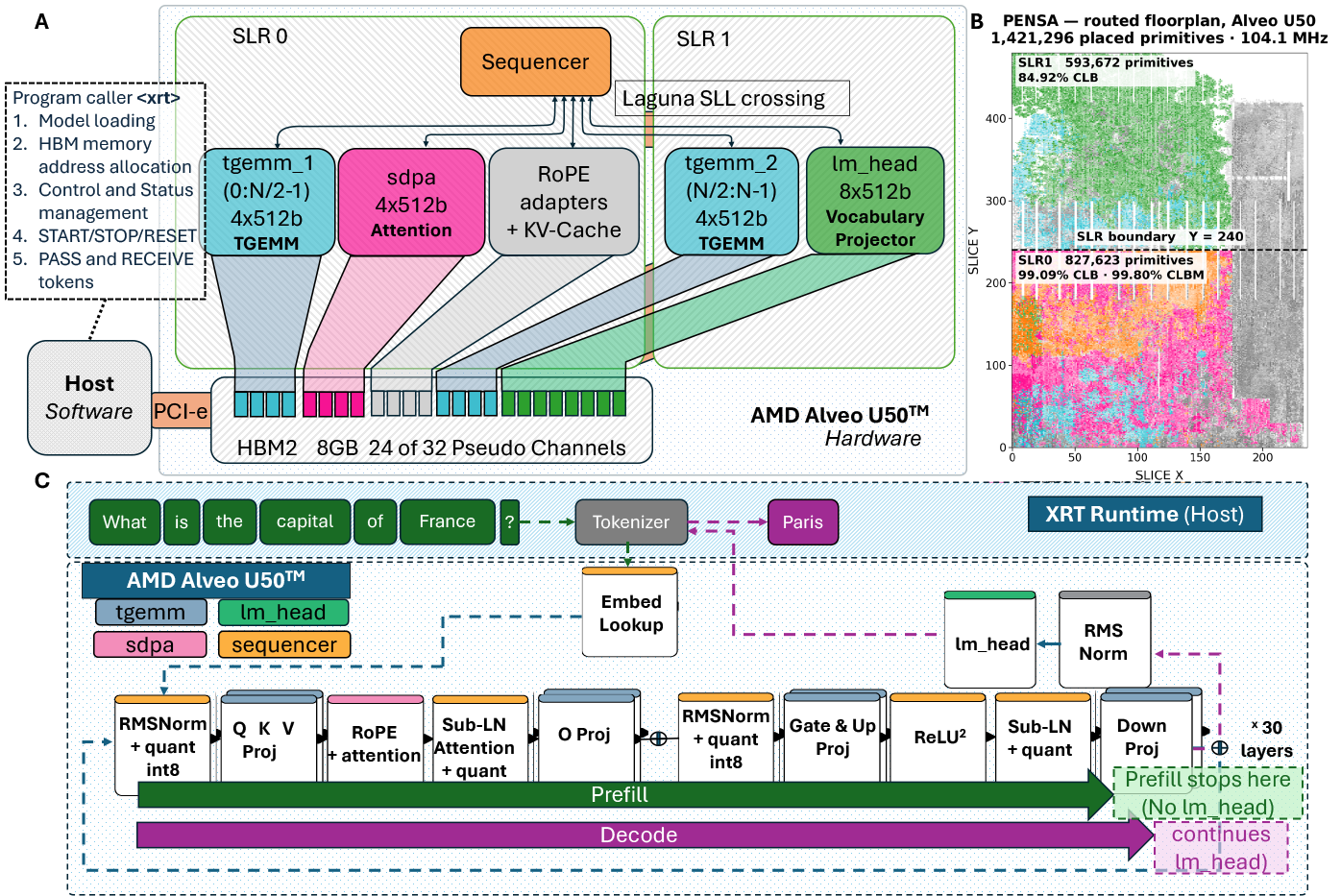}
\caption{\textsc{PENSA} on the AMD Alveo U50\texttrademark:
(A) Penta-Unit with assigned HBM interfaces; (B) routed floorplan; and (C) resident BitNet~b1.58~2B4T token-execution dataflow.}
\label{fig:arch}

\vspace{-5mm}
\end{figure}

\section{\textsc{PENSA} Architecture}

\subsection{Penta-Unit Spatial Accelerator Dataflow}

Figure~\ref{fig:arch}A shows the Penta-Unit, \textsc{PENSA}'s five persistent compute units: the {\tt Sequencer}, two ternary matrix engines ({\tt tgemm\_1} in {\tt SLR~0} and {\tt tgemm\_2} in {\tt SLR~1}), the attention engine ({\tt sdpa}), and the vocabulary projector ({\tt lm\_head}). After loading the bitstream and model into HBM and starting the units, the {\tt Sequencer} controls prefill and decode autonomously. Independently assigned HBM pseudo-channels supply the dominant data streams. The routed AMD~Alveo~U50\texttrademark~floorplan is shown in Figure~\ref{fig:arch}B.

Figure~\ref{fig:arch}C shows the resident execution path. The {\tt Sequencer} performs {\tt Embed Lookup}, normalization, and activation quantization across all 30 transformer layers. Each layer's dual TGEMMs execute the seven ternary projections (Q, K, V, O, Gate, Up, and Down). {\tt sdpa} applies RoPE, updates the HBM-resident K/V cache, and evaluates grouped-query attention using streaming softmax, without materializing the \(T\times T\) score matrix. After the O and Gate/Up/Down paths, ReLU\(^{2}\), and residual updates, final normalization feeds the BF16 {\tt lm\_head} with fused on-chip argmax. Only the selected token ID returns to the {\tt Sequencer}, keeping execution and autoregressive feedback on the FPGA.
\begin{figure}[ht]
\centering
\includegraphics[width=0.75\linewidth]{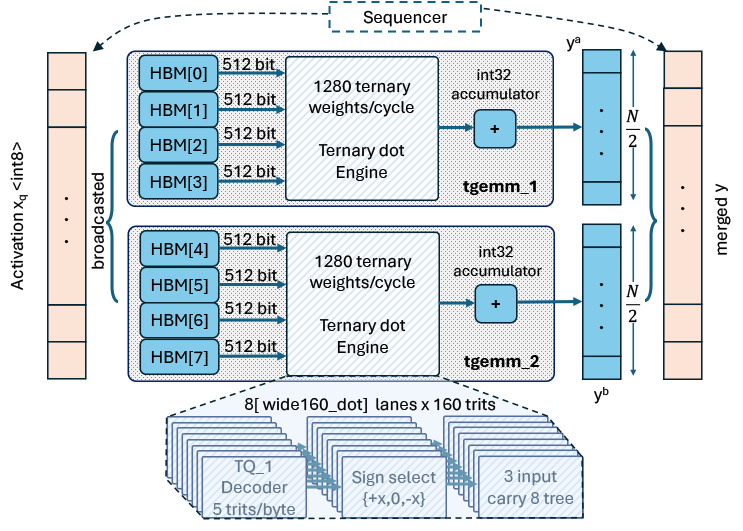}
\caption{Dual-TGEMM datapath. A shared INT8 activation is broadcast to two row-partitioned engines, each supplied by four HBM pseudo-channels. Eight {\tt wide160\_dot} lanes per TGEMM process 1,280 packed ternary weights per cycle.}
\label{fig:dual-tgemm}

\vspace{-5mm}
\end{figure}
\subsection{Dual Ternary Matrix Engines}

Figure~\ref{fig:dual-tgemm} shows the dual-TGEMM organization. The {\tt Sequencer} quantizes each normalized activation to INT8 once and
broadcasts the resulting vector \(\mathbf{x}_q\), together with its scale, to both persistent TGEMMs. For \(\mathbf{W}\in\{-1,0,+1\}^{N\times K}\), ternary multiplication reduces to sign-controlled selection:
\begin{equation}
y_i=\sum_{j:W_{ij}=+1}x_{q,j}
    -\sum_{j:W_{ij}=-1}x_{q,j}.
\label{eq:ternary-dot}
\end{equation}
Thus, the integer dot-product datapath requires no general multiplier or DSP-based MAC. Output rows are divided evenly between the engines,
\begin{equation}
\mathbf{W}=
\begin{bmatrix}
\mathbf{W}^{(a)}\\
\mathbf{W}^{(b)}
\end{bmatrix},
\quad
\mathbf{y}^{(a)}=\mathbf{W}^{(a)}\mathbf{x}_q,\quad
\mathbf{y}^{(b)}=\mathbf{W}^{(b)}\mathbf{x}_q,\quad
\mathbf{y}=
\begin{bmatrix}
\mathbf{y}^{(a)}\\
\mathbf{y}^{(b)}
\end{bmatrix},
\quad
\mathbf{W}^{(a,b)}
\in\{-1,0,+1\}^{(N/2)\times K}.
\label{eq:dual-tgemm}
\end{equation}
{\tt tgemm\_1} consumes weight stripes from HBM pseudo-channels 0--3 and {\tt tgemm\_2} from 4--7; the {\tt Sequencer} merges their complementary output rows. Both engines therefore cooperate on the same projection rather than processing independent sequences.

\paragraph{Direct packed-weight datapath.}
Weights remain in the model's TQ1\_0 base-3 representation, which packs five ternary values per byte, corresponding to a 1.6-bit/weight packed
payload~\cite{bitnetcpp}. Each 512-bit AXI beat comprises two 256-bit TQ1\_0 blocks, each encoding 160 ternary weights. Four interfaces therefore supply eight blocks, or 1,280 ternary weights, to each TGEMM per cycle. A compile-time 256-entry decode ROM maps each byte
to five ternary controls, which directly select \(0\), \(+x_{q,j}\), or \(-x_{q,j}\); no activation-dependent lookup table or multiplier is required.

Each {\tt wide160\_dot} pads 160 selected INT8 terms with two zeros.
Three 3:1 reduction stages produce 54 signed 11-bit, 18 signed
13-bit, and six signed 15-bit partial sums. A final reduction adds
the six partial sums into a signed 17-bit dot result, followed by
INT32 accumulation across input tiles. Activations are clamped to
\([-127,127]\), so ternary negation remains representable in INT8.
The datapath sustains II=1, with eight lanes per TGEMM processing
1,280 ternary weights per cycle. Relative to the preceding linear
reduction, it reduces LUT and FF use by 12\% and 76\%, respectively.

\paragraph{Two-stream interleaving.}
The TGEMMs are stateless projection servers: both always cooperate on one slot and projection, while the {\tt Sequencer} temporally interleaves two independent sequence slots across the persistent pipeline. For example, while {\tt lm\_head} processes one slot, the other can advance through transformer layers using both TGEMMs. Each slot retains independent token and K/V-cache state, improving cross-unit occupancy without changing the model computation.
\begin{figure}[ht]
\centering
\includegraphics[width=0.95\linewidth]
{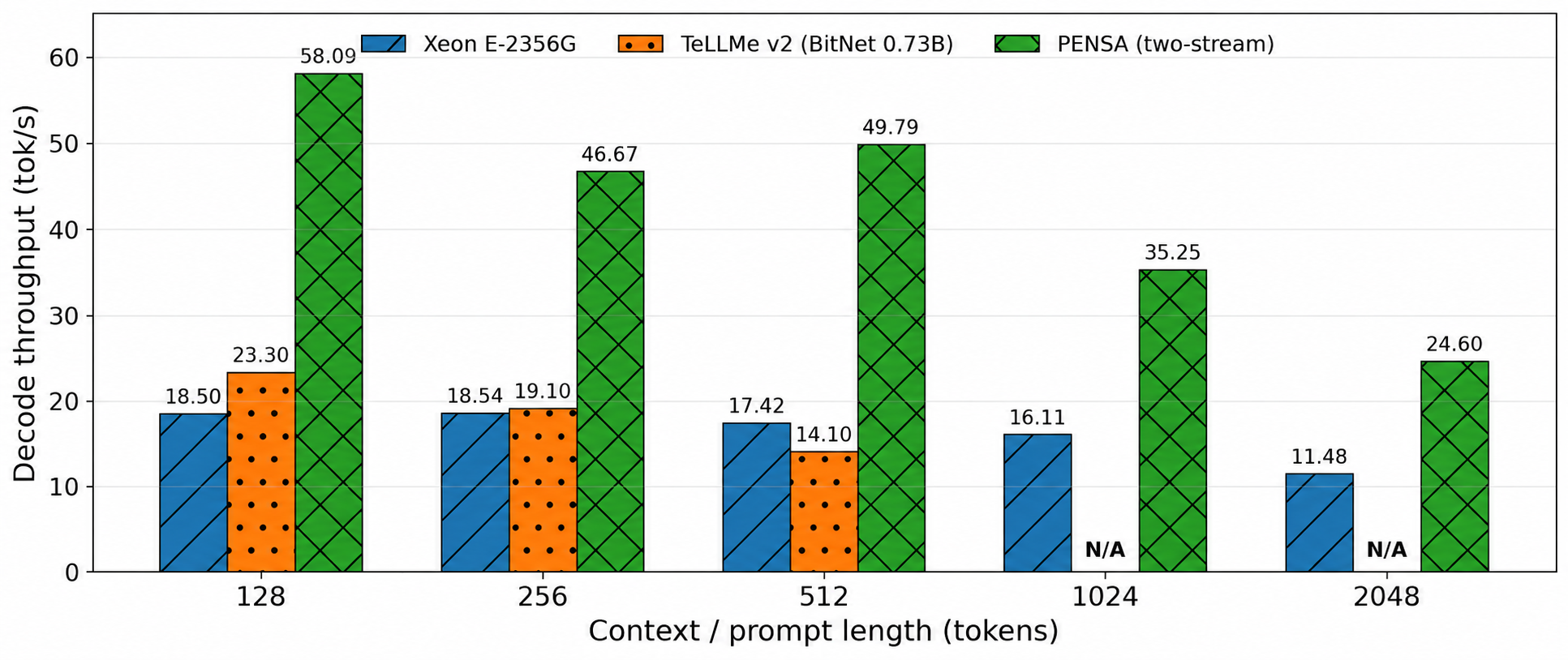}
\caption{Decode throughput of \textsc{PENSA}, the same-model Xeon E-2356G baseline, and TeLLMe~v2~\cite{qiao2026tellmev2}. \textsc{PENSA} and Xeon evaluate BitNet~b1.58~2B4T at decode contexts 128--2048. TeLLMe~v2 evaluates on BitNet~0.73B at [prompt,generation] configurations [128,128], [256,128], and [512,128], with longer lengths unreported.}
\label{fig:platform-decode}

\vspace{-5mm}
\end{figure}

\section{Evaluation}

Figure~\ref{fig:platform-decode} compares decode throughput across context lengths. At the achieved post-route frequency of 104.1~MHz, limited by the platform HBM/SLR read-response path rather than a \textsc{PENSA} compute unit (Appendix~\ref{app:model}), two-stream interleaving achieves 58.09~tok/s at context 128 and 24.60~tok/s at 2048, yielding 2.14--3.14$\times$ the throughput of the same-model Xeon E-2356G~baseline. Interleaving improves aggregate throughput by 1.37--1.71$\times$ over a single-stream execution through persistent unit overlap. However, it does not amortize the {\tt lm\_head} weight stream across sequences. Table~\ref{tab:ternary-sota}~provides cross-model architectural positioning. While TerEffic\cite{yin2025tereffic} has a higher reported throughput, it reflects both cross-platform/model differences and compute--memory co-design that reduces weight traffic. PENSA instead targets the unmodified BitNet~b1.58~2B4T topology, including its 328.3M-parameter BF16 vocabulary projection that moves approximately 656.6~MB per decoded token versus 416.8~MB of packed ternary weights, corresponding to transfer times of 12.3 and 7.8~ms, respectively (Appendix~\ref{app:model}). Prefill increases from 6.55~s at 128~tokens to 109.18~s at 2048~tokens (see Appendix~\ref{app:performance}).

At context 2048, the context-dependent term contributes approximately 60\% of modeled decode latency, with attention traversal accounting for 97.3\% of its fitted per-key component (Appendix~\ref{app:performance}). A 2,400-token two-stream run at 46.18~tok/s draws \(33.46\pm0.24\)~W corresponding to 0.725~J/token, or 0.405~J/token above the 14.75~W idle baseline (Appendix~\ref{app:power}). Table~\ref{tab:accuracy} reports FPGA model-fidelity results, which differ from the published task accuracies by 0.75--1.87 percentage points and PPL by only 0.07. Evaluation settings, uncertainty definitions, and the teacher-forced scoring path are given in Appendix~\ref{app:evaluation} and \ref{app:PPL}.
\begin{table}[ht]
\centering
\caption{Architectural positioning of ternary-LLM FPGA accelerators;
cross-model throughput is not directly comparable.}
\label{tab:ternary-sota}
\small
\setlength{\tabcolsep}{3pt}
\begin{tabular}{@{}l l l c p{5.5cm}@{}}
\toprule
Design & Model & FPGA & Reported tok/s & Distinguishing feature \\
\midrule

TeLLMe~\cite{qiao2025tellme}
 & BitNet 0.7B
 & KV260
 & 9.51
 & Lookup-based TMM \\

TeLLMe v2~\cite{qiao2026tellmev2}
 & BitNet 0.73B
 & KV260
 & 14.1--23.3
 & TLMM, on-chip weight buffering \\

TerEffic~\cite{yin2025tereffic}
 & Ternary 2.7B
 & Alveo U280
 & 521
 & HBM-assisted compute--memory co-design \\

TENET~\cite{huang2025tenet}
 & Ternary LLM
 & Stratix 10 MX
 & --
 & N:M sparsity, weight decompression \\

\textsc{PENSA}
 & BitNet 2B4T
 & Alveo U50
 & 24.60--58.09
 & Direct TQ1\_0, resident execution \\

\bottomrule
\end{tabular}

\vspace{-5mm}
\end{table}
\begin{table}[ht]
\centering
\caption{Model fidelity vs. published BitNet~b1.58~2B~\cite{ma2025bitnet}; accuracy differences are pp,~PPL absolute.}
\label{tab:accuracy}
\small
\setlength{\tabcolsep}{3pt}
\begin{tabular}{@{}lccccc@{}}
\toprule
 & ARC-E & ARC-C & MMLU & GSM8K & PPL (WikiText-2) \\
\midrule
Software Reference
 & 74.79 & 49.91 & 53.17 & 58.38 & 14.59 \\
\textsc{PENSA}
 & $73.57{\pm}.90$ & $48.04{\pm}1.46$
 & $52.39{\pm}.69$ & $57.63{\pm}0.82$ & 14.52 \\
$\Delta$
 & -1.22 & -1.87 & -0.78 & -0.75 & -0.07 \\
\bottomrule
\end{tabular}
\vspace{-5mm}

\end{table}

\section{Conclusion}
\textsc{PENSA} demonstrates end-to-end BitNet~b1.58~2B4T inference on an AMD Alveo U50\texttrademark~ without modifying the model topology. At 104.1~MHz, two-stream interleaving reaches 58.09~tok/s and exceeds the same-model Xeon baseline across all evaluated contexts, while preserving the model fidelity. Native ternary arithmetic shifts the limiting factors from multiplication toward packed-weight supply, reduction, persistent-unit occupancy, attention, and vocabulary projection.

\begin{ack}
The authors gratefully acknowledge support from the Australian Economic Accelerator Ignite Grant IG240100745 and the AMD University Program.
\end{ack}

\bibliographystyle{unsrt}
\bibliography{references}

\appendix

\section{Model and HBM Mapping}
\label{app:model}

BitNet~b1.58~2B4T configuration used by \textsc{PENSA} contains 30 transformer layers with hidden dimension \(C=2560\), feed-forward dimension \(C_{\mathrm{ff}}=6912\), grouped-query K/V dimension \(C_{\mathrm{kv}}=640\), and vocabulary size \(V=128256\). Each layer contains seven ternary projections: Q, K, V, O, Gate, Up, and Down. Across the transformer stack, the TGEMMs process 2.084B ternary linear weights, of which 1.593B (76.4\%) belong to the feed-forward projections and 0.492B (23.6\%) to the attention projections. The separate BF16 vocabulary projection contains 328.3M parameters.

Table~\ref{tab:hbm-map} summarizes the principal HBM ownership. Dominant weight streams use independent pseudo-channels to avoid arbitration between concurrent consumers.

\begin{table}[h]
\centering
\small
\caption{Principal HBM ownership in \textsc{PENSA}.}
\label{tab:hbm-map}
\begin{tabular}{@{}lll@{}}
\toprule
Unit & HBM PCs & Contents \\
\midrule
TGEMM A & 0--3  & packed TQ1\_0 weights \\
TGEMM B & 4--7  & packed TQ1\_0 weights \\
{\tt lm\_head} & 8--15 & BF16 vocabulary weights \\
Sequencer & 19--24 & model data, layout, token state \\
{\tt sdpa} & 26--29 & K/V cache and RoPE tables \\
\bottomrule
\end{tabular}
\end{table}

Twenty-six of the 32 pseudo-channels carry data in the primary mapping. The remaining channels are unused or correspond to NULL-bound interfaces sharing existing bundles. The 2.084B ternary weights correspond to approximately 416.8~MB of packed payload per transformer pass, whereas the 328.3M-parameter BF16 vocabulary projection requires approximately 656.6~MB per decoded token. At 104.1~MHz, eight 512-bit interfaces provide a nominal aggregate issue rate of 53.3~GB/s, giving payload-only traffic floors of approximately 7.8~ms for the ternary weights and 12.3~ms for the vocabulary projection. Thus, despite its smaller parameter count, the unquantized vocabulary projection has the larger per-token weight stream and constitutes an important optimization target. The measured pseudo-channel ceiling is 13.1~GB/s. The 104.1~MHz operating point is the achieved post-route frequency. The critical path in the platform HBM infrastructure ({\tt hmss\_0}) is the AXI read-response path crossing an SLR boundary through Laguna registers, rather than in
a \textsc{PENSA} compute unit. Consequently, weight traffic, request rate, recurrence, consumer occupancy, and implementation timing can limit execution before the physical HBM-channel ceiling is reached.

\section{Host Execution and Reproducibility}
\label{app:host}

\textsc{PENSA} uses a resident hardware--software execution model,
following the broader FPGA--host co-design principle of separating
software orchestration from hardware-resident dataflow.
The host loads the bitstream and model, allocates HBM buffers, configures the five persistent compute units, and starts each unit once. Prefill and autoregressive decode then execute on device; the host may poll a progress counter for reporting but performs no per-token computation or kernel launch. Generated token IDs are transferred to the host only after execution, and no AXI4-Stream traffic leaves the device during
generation. Tokenization and final detokenization remain CPU-side operations.

The implementation uses AMD Vitis 2025.2 and the corresponding Xilinx Runtime (XRT). XRT performs device configuration, HBM buffer allocation and binding, kernel argument setup, persistent kernel launch, status polling, and final token ID retrieval. Model and auxiliary buffers are allocated before kernel start; the final normalization scale is aligned to the 32-element boundary required by the kernel's wide memory accesses.

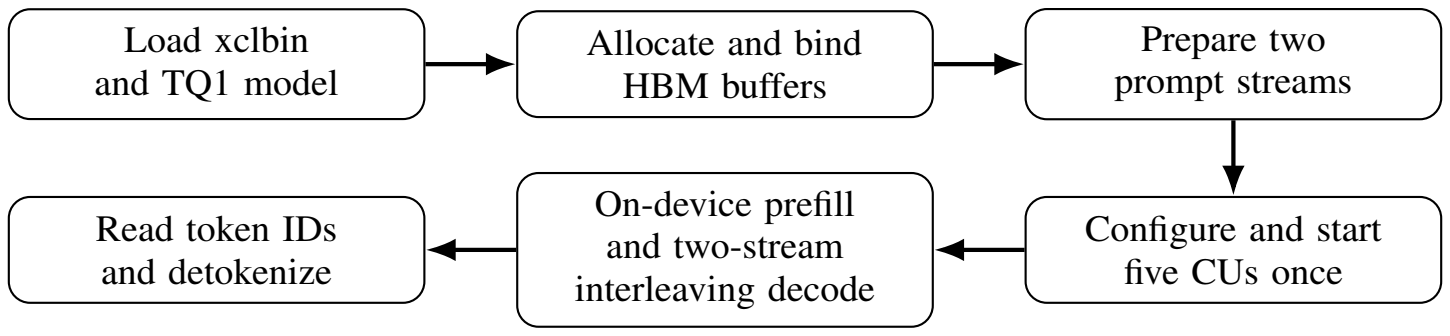
\begin{figure}[h]
\centering
\begin{tikzpicture}[
  node distance=5mm and 6mm,
  box/.style={draw,rounded corners,align=center,minimum height=7mm,
              text width=0.18\linewidth,font=\scriptsize},
  arr/.style={-{Latex[length=2mm]},thick}
]
\node[box] (load) {Load xclbin\\and TQ1 model};
\node[box,right=of load] (alloc) {Allocate and bind\\HBM buffers};
\node[box,right=of alloc] (seed) {Prepare two\\prompt streams};
\node[box,below=of seed] (start) {Configure and start\\five CUs once};
\node[box,left=of start] (run) {On-device prefill\\and two-stream interleaving decode};
\node[box,left=of run] (read) {Read token IDs\\and detokenize};

\draw[arr] (load)--(alloc);
\draw[arr] (alloc)--(seed);
\draw[arr] (seed)--(start);
\draw[arr] (start)--(run);
\draw[arr] (run)--(read);
\end{tikzpicture}
\caption{Host execution flow. Kernel launch occurs once; prefill, decode, and autoregressive token feedback remain resident on the FPGA.}
\label{fig:hostflow}
\end{figure}

For deterministic reference comparisons, greedy decoding is used, and the squared-ReLU feed-forward activation is enabled. Two-stream interleaving uses two equal-length prompts because both sequence slots advance under a shared token-loop position while retaining independent token and K/V state.

Generations are checked for prefix consistency across increasing generation lengths, and two-stream interleaving is exercised with different prompts to verify independent sequence state. The host executable and xclbin are treated as a matched pair because the kernel argument layout is fixed by the compiled design.

\section{TGEMM Microarchitecture}
\label{app:tgemm}

TQ1\_0 stores five ternary weights per byte, yielding a 1.6-bit/weight packed payload. A 256-bit memory word therefore contains 32 encoded bytes representing 160 ternary weights. Each byte addresses a compile-time 256-entry decode ROM that produces
five ternary controls. Activations are clamped to \([-127,127]\);
the controls therefore select \(0\), \(+x_{q,j}\), or \(-x_{q,j}\)
without exceeding the signed INT8 range. The accompanying FP32
activation scale is applied outside the integer datapath.

Each {\tt wide160\_dot} pads its 160 selected INT8 terms with two
zeros. The reduction forms 54 signed 11-bit sums, then 18 signed
13-bit sums, then six signed 15-bit sums. The six partial sums are added in the final reduction into a signed 17-bit dot result before INT32 accumulation
across input tiles. The maximum magnitude of a valid 160-term dot
product is \(160\times127=20{,}320\), which fits in signed 16 bits.
The datapath sustains II=1. Eight lanes operate in each TGEMM,
corresponding to 1,280 ternary weights per cycle per engine. TGEMM A and
TGEMM B use four independent HBM pseudo-channels each and process
complementary output-row partitions of the same projection. Post-route implementation uses 509,234 LUTs (58.5\%), 706,614 FFs
(40.6\%), 931.5 BRAMs (69.3\%), 274 URAMs (42.8\%), and 2,714 DSPs
(45.6\%) of the U50. The DSP total is whole-design utilization and
includes non-TGEMM operators; the ternary dot datapaths themselves use
no DSP-based multiplication. Relative to the preceding linear
reduction, the carry-oriented tree reduces LUT and FF use by 12\% and
76\%, respectively.

\begin{table}[h]
\centering
\small
\caption{Post-route resource utilization of the persistent kernels.}
\label{tab:unit-resources}
\begin{tabular}{@{}lrrrrr@{}}
\toprule
Kernel & LUT & FF & BRAM & URAM & DSP \\
\midrule
{\tt Sequencer} & 68{,}592  & 79{,}221  & 111 & 28 & 216 \\
{\tt tgemm\_1}  & 29{,}028  & 26{,}069  & 180 & 82 & 27 \\
{\tt tgemm\_2}  & 28{,}983  & 26{,}070  & 180 & 82 & 27 \\
{\tt sdpa}      & 104{,}023 & 145{,}193 & 122 & 60 & 1{,}111 \\
{\tt lm\_head}  & 120{,}791 & 176{,}039 & 132 & 8  & 1{,}329 \\
\midrule
Total (kernels)    & 351{,}417 & 452{,}592 & 725 & 260 & 2{,}710 \\
\bottomrule
\end{tabular}
\end{table}

The two TGEMMs have nearly identical utilization, consistent with
their symmetric row partitioning. Together they account for only
54 of the 2,710 kernel-level DSPs; the ternary dot datapaths themselves
use no DSP-based multiplication, while the reported TGEMM DSPs arise
from surrounding non-ternary arithmetic. In contrast, \texttt{sdpa}
and \texttt{lm\_head} account for 2,440 DSPs (90.0\% of the kernel
total), illustrating that native ternary execution shifts arithmetic
resource pressure toward the remaining mixed-precision operators.
Kernel totals exclude platform, HBM, and interconnect infrastructure
and therefore do not equal the whole-device utilization.

\section{Additional Performance Results}
\label{app:performance}

Table~\ref{tab:batch2} reports single-stream and two-stream interleaving decode throughput at 104.1~MHz. Two-stream interleaving raises aggregate throughput by 1.37--1.71$\times$ across the evaluated contexts by overlapping independent sequence slots within the persistent units; both TGEMMs continue to cooperate on one projection at a time.

\begin{table}[h]
\centering
\small
\caption{Single-stream and two-stream interleaving decode throughput at 104.1~MHz.}
\label{tab:batch2}
\begin{tabular}{@{}rrrr@{}}
\toprule
Context & Single (tok/s) & Two-stream (tok/s) & Gain \\
\midrule
128  & 37.15 & 58.09 & 1.56$\times$ \\
256  & 33.49 & 46.67 & 1.39$\times$ \\
512  & 29.07 & 49.79 & 1.71$\times$ \\
1024 & 25.64 & 35.25 & 1.37$\times$ \\
2048 & 17.92 & 24.60 & 1.37$\times$ \\
\bottomrule
\end{tabular}
\end{table}

Measured prefill times at sequence lengths 128, 256, 512, 1024, and 2048 are 6.55, 7.80, 21.30, 42.42, and 109.18~s, respectively.
The measured average interval per generated token, computed from aggregate
two-stream throughput, is approximated by
\begin{equation}
t_{\mathrm{agg}}(T)
\simeq
16.15~\mathrm{ms}
+
(0.01186~\mathrm{ms/key})T.
\end{equation}
Here \(T\) is the active context length. This aggregate output interval
is not the latency of an individual sequence's token. The fit gives \(R^2=0.972\), RMSE 1.40~ms, and a maximum residual
of 2.25~ms (10.5\%) at \(T=256\). Measured throughput is locally
non-monotonic between contexts 256 and 512. The present measurements
do not isolate this deviation, so the linear model is used only to
characterize the overall context-dependent trend. At \(T=2048\), the context-dependent component contributes approximately 60\% of modeled latency. Operator-level timing attributes approximately 97.3\% of the fitted per-key term to attention traversal.

\section{Power Measurement}
\label{app:power}

Power is sampled using \texttt{xrt-smi} JSON telemetry during a
2,400-token two-stream decode run completing in 51.97~s
(46.18~tok/s aggregate). Table~\ref{tab:power} reports idle and active
rail power together with dynamic power, defined as active minus idle.
Energy per token is computed from the corresponding power divided by
measured aggregate throughput.

\begin{table}[h]
\centering
\small
\caption{Power and energy during two-stream decode.}
\label{tab:power}
\begin{tabular}{@{}lrrrr@{}}
\toprule
Rail & Idle (W) & Active (W) & Dynamic (W) & Energy/token \\
\midrule
Board
 & $14.75\pm0.02$ & $33.46\pm0.24$ & 18.70
 & 0.725 active / 0.405 dynamic \\
VCCINT
 & $6.37\pm0.03$ & $15.51\pm0.74$ & 9.14
 & 0.336 active / 0.198 dynamic \\
VCCINT\_IO
 & $0.44\pm0.02$ & $3.73\pm0.06$ & 3.29
 & 0.081 active / 0.071 dynamic \\
\bottomrule
\end{tabular}
\end{table}

The VCCINT rail contributes 15.51~W active and 9.14~W above idle,
corresponding to 0.336 and 0.198~J/token, respectively. These values
represent FPGA-core rail consumption rather than TGEMM-only power.
VCCINT also exhibits greater variation than board-level power,
consistent with phase-dependent activity within the FPGA fabric.

\section{Evaluation Methodology}
\label{app:evaluation}

FPGA likelihood evaluation is controlled by the host and uses a teacher-forced scoring mode in which the host supplies the token sequence rather than feeding back the on-chip argmax. At a host-selected position, {\tt lm\_head} returns logits for requested token indices together with sufficient statistics.
For every vocabulary partition~$p \in \{0,\ldots,7\}$, let $z_{p,i}$ denote the logit of vocabulary entry $i$ in partition $p$. The FPGA returns the partition maximum and shifted exponential sum, 

\[
m_p=\max_i z_{p,i},\qquad
s_p=\sum_i \exp(z_{p,i}-m_p).
\]
Here, $p$ indexes the eight vocabulary partitions and $i$ indexes vocabulary entries within a partition. The host combines these summaries by computing
\[
M=\max_p m_p,\qquad
\log Z=M+\log\sum_p s_p e^{m_p-M},
\] 
where $M$ is the global maximum and $Z$ is the full-vocabulary softmax normalization constant. For a requested target token $t$, whose logit $z_t$ is returned explicitly by the FPGA, its log-likelihood is then 
\[
\log P(t)=z_t-\log Z.
\]

Thus, exact token likelihoods can be reconstructed without transferring the complete vocabulary distribution. Only the requested token logits and two summary statistics ($m_p, s_p$) from each of the eight partitions are returned, requiring approximately 80 bytes per scored position. 

A custom EleutherAI \texttt{lm-evaluation-harness} v0.4.12 adapter
uses this scoring path for likelihood-based tasks and the native
autoregressive path for generation. MMLU uses \texttt{cais/mmlu}
version~1.0 with the test split, five-shot examples from the dev split,
the deterministic \texttt{first\_n} sampler, and accuracy as the
metric. The same five dev examples are therefore shared by all test
items within a subject and their K/V state is reused. Because the four
MMLU alternatives are single-token choices, their logits are obtained
from one scored position.

GSM8K uses \texttt{generate\_until} with greedy decoding
(\texttt{do\_sample=false}, temperature~0), the benchmark stop strings,
and the \texttt{flexible-extract} filter. The reported FPGA result uses
four-shot prompting and achieves \(0.5763\pm0.0082\) exact-match
accuracy. Reported $\pm$ values are harness-reported standard errors
across evaluation items within a single evaluation pass, rather than
variation across repeated runs.
ARC-Easy and ARC-Challenge use

{\tt lm-evaluation-harness} v0.4.12 task version~1.0 in the
0-shot setting with \texttt{multiple\_choice} output. Each candidate's
full answer text is scored as a continuation of
``Question: \{question\}\textbackslash n~Answer:'', and the
highest-likelihood candidate is selected. Both \texttt{acc} and
byte-length-normalized \texttt{acc\_norm} are recorded; comparisons
with published results use the same shot count and metric variant.
\begin{figure}[t]
\centering
\includegraphics[width=0.85\linewidth]{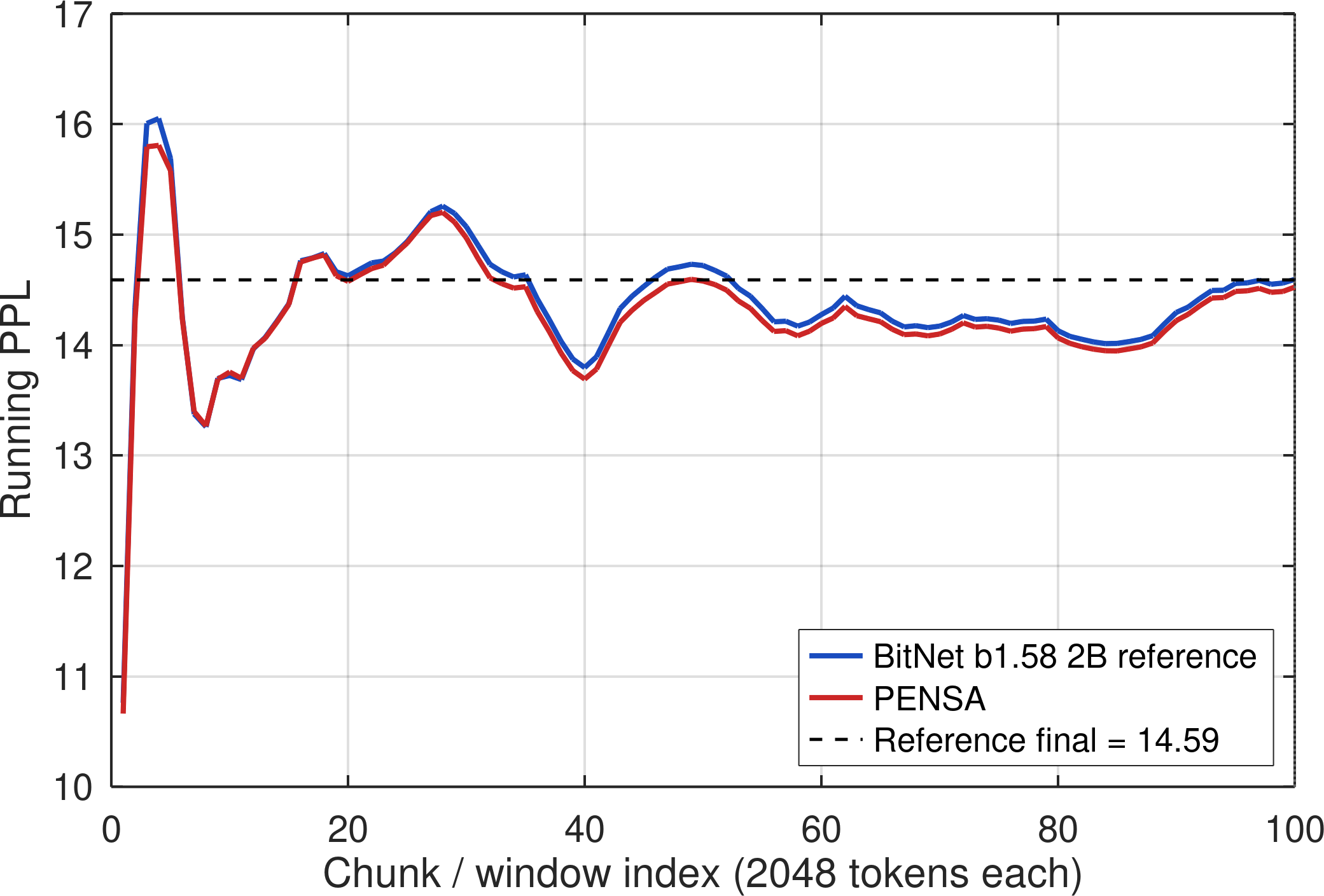}
\caption{Running perplexity of \textsc{PENSA} (red) and BitNet~b1.58~2B reference (blue) as a function of chunk index. The reference PPL value reported in Table~\ref{tab:accuracy} is indicated by the dashed line.}
\label{fig:PPL}
\end{figure}

\section{Perplexity Computation}
\label{app:PPL}

Perplexity (PPL) is a standard intrinsic measure of LLM quality that evaluates the model's ability to predict every token in a text corpus. For a sequence of $N$ tokens, $x_1,\ldots,x_N$, an autoregressive model assigns each token the conditional probability $P(x_i\mid x_1,\ldots,x_{i-1})$. Perplexity is the exponential of the average negative log-likelihood:
\begin{equation}
\mathrm{PPL}
=
\exp\!\left[
-\frac{1}{N-1}
\sum_{i=2}^{N}
\log P(x_i\mid x_1,\ldots,x_{i-1})
\right].
\end{equation}
Lower PPL indicates more accurate prediction of the observed sequence; a perfect model would have $\mathrm{PPL}=1$.

Each conditional $\log P(x_i\mid\cdot)$ is obtained through the
teacher-forced scoring path described in Appendix~\ref{app:evaluation}.
The host supplies ground-truth tokens instead of feeding back the
on-chip argmax. For each scored position, \texttt{lm\_head} returns
the partition statistics $(m_p,s_p)$ and the requested target logit,
allowing the host to reconstruct $\log P(x_i)=z_{x_i}-\log Z$.
One device invocation advances through the tokens of a context window,
retains their K/V state, and returns the likelihood statistics for
scored positions. This avoids a separate host launch at each position
but does not eliminate context-dependent attention work.

Because on-chip state is reset at each context-window boundary, a chunked evaluation conditions each prediction on only the tokens within its own window, not the full corpus prefix. For a corpus of $N$ tokens partitioned into $K=\lceil N/W\rceil$ windows of length $W$ (the final window may be shorter), with window $k$ spanning tokens $x_{b_k+1},\ldots,x_{e_k}$ for $b_k=(k-1)W$ and $e_k=\min(kW,N)$, the
first token of each window has no in-window context and is excluded from scoring, giving
\begin{equation}
\mathrm{PPL}
=
\exp\!\left[
-\frac{1}{N-K}
\sum_{k=1}^{K}\ \sum_{j=b_k+2}^{e_k}
\log P\bigl(x_j \mid x_{b_k+1},\ldots,x_{j-1}\bigr)
\right],
\end{equation}
which reduces to the single-sequence expression above when $K=1$ ($W\geq N$). Equivalently, log-likelihoods and scored-token counts are accumulated across all windows before the single final exponentiation (not averaged per-window) since the latter is not equivalent for windows of unequal length and is sensitive to how loss is weighted across them.

PPL is sensitive to the choice of context window $W$ and stride $S\leq W$:~small windows inflate PPL because every token near a window's start is scored with truncated context, and this effect is independent of any model or hardware. In the results reported here, both the reference model and \textsc{PENSA} were evaluated on the entire WikiText-2 test corpus with $W=S=2048$, i.e.~non-overlapping windows at the model's full context length, matching \texttt{llama.cpp}'s default (\texttt{--ppl-stride} unset) chunking behavior ($\mathtt{start}=i\cdot W$, $\mathtt{end}=\mathtt{start}+W$, cache cleared between chunks).

The customized PPL procedure enables comparison of per-chunk PPL dynamics against the reference. Over 100 representative measured chunks (Figure~\ref{fig:PPL}), \textsc{PENSA} (red) closely tracks the published BitNet~b1.58~2B reference (blue) (MAE = 0.0771, RMSE = 0.0878, Pearson $r = 0.997$, mean absolute relative difference = 0.53\%), with a maximum single-chunk discrepancy of 1.51\%. Running PPL oscillates sharply over the first 40 chunks before settling near their respective final values of 14.52 for PENSA and 14.59 for the reference. This behavior reflects the cumulative-average nature of PPL and the heterogeneous composition of the WikiText-2 test file, which concatenates distinct Wikipedia articles rather than forming a statistically uniform text stream. As the evaluation window moves between articles, per-chunk PPL therefore rises and falls with the underlying content rather than merely reflecting noise. The near-identical agreement between \textsc{PENSA}'s PPL curve and BitNet~b1.58~2B reference, particularly through this oscillatory region, provides evidence that these variations are reproducible properties of the text and model rather than implementation artifacts.
\newpage
\begin{itemize}
    \item {\bf Claims}
    \item[] Question: Do the main claims made in the abstract and introduction accurately reflect the paper's contributions and scope?
    \item[] Answer: \answerYes{}
    \item[] Justification: The abstract and Introduction state the scope of
    \textsc{PENSA} as an end-to-end FPGA implementation of
    BitNet~b1.58~2B4T and distinguish same-model CPU comparisons from
    cross-model architectural comparisons. The reported throughput,
    fidelity, and bottleneck observations are supported by the
    Evaluation and appendices.

\item {\bf Limitations}
    \item[] Question: Does the paper discuss the limitations of the work performed by the authors?
    \item[] Answer: \answerYes{}
    \item[] Justification: The Evaluation and Appendix~D report the strong context dependence of decode latency and the substantially increasing
    prefill latency. Cross-model accelerator results are explicitly presented as architectural context rather than direct performance
    comparisons.

\item {\bf Theory assumptions and proofs}
    \item[] Question: For each theoretical result, does the paper provide the full set of assumptions and a complete (and correct) proof?
    \item[] Answer: \answerNA{}
    \item[] Justification: The paper presents an accelerator architecture
    and empirical evaluation and does not claim new theoretical results
    requiring formal proofs.

\item {\bf Experimental result reproducibility}
    \item[] Question: Does the paper fully disclose all the information needed to reproduce the main experimental results of the paper to the extent that it affects the main claims and/or conclusions of the paper (regardless of whether the code and data are provided or not)?
    \item[] Answer: \answerYes{}
    \item[] Justification: Sections~2--3 and Appendices~A--E describe the
    model dimensions, ternary representation, TGEMM organization, HBM
    mapping, host execution, implementation resources, evaluation
    methodology, and performance measurements needed to reproduce or
    independently verify the reported results.

\item {\bf Open access to data and code}
    \item[] Question: Does the paper provide open access to the data and code, with sufficient instructions to faithfully reproduce the main experimental results, as described in supplemental material?
    \item[] Answer: \answerNo{}
    \item[] Justification: The paper provides implementation and
    reproducibility details, but the complete FPGA implementation and
    experiment scripts are not currently released as an anonymized
    open-access artifact.

\item {\bf Experimental setting/details}
    \item[] Question: Does the paper specify all the training and test details (e.g., data splits, hyperparameters, how they were chosen, type of optimizer) necessary to understand the results?
    \item[] Answer: \answerYes{}
    \item[] Justification: This work evaluates inference rather than
    training. The model, hardware configuration, operating frequency,
    context lengths, execution modes, CPU baseline, and evaluation
    methodology are specified in the Evaluation and appendices.

\item {\bf Experiment statistical significance}
    \item[] Question: Does the paper report error bars suitably and correctly defined or other appropriate information about the statistical significance of the experiments?
    \item[] Answer: \answerYes{}
    \item[] Justification: Table~2 reports uncertainty for the downstream
    evaluation tasks, with its definition and evaluation procedure given
    in Appendix~E. Appendix~D additionally reports $R^2$, RMSE, and the
    maximum residual for the decode-latency fit.

    \item {\bf New assets}
    \item[] Question: Are new assets introduced in the paper well documented and is the documentation provided alongside the assets?
    \item[] Answer: \answerNA{}
    \item[] Justification: The submission does not release a new dataset,
    pretrained model, or public software artifact.

\item {\bf Crowdsourcing and research with human subjects}
    \item[] Question: For crowdsourcing experiments and research with human subjects, does the paper include the full text of instructions given to participants and screenshots, if applicable, as well as details about compensation (if any)?
    \item[] Answer: \answerNA{}
    \item[] Justification: The research does not involve crowdsourcing or
    human subjects.

\item {\bf Institutional review board (IRB) approvals or equivalent for research with human subjects}
    \item[] Question: Does the paper describe potential risks incurred by study participants, whether such risks were disclosed to the subjects, and whether Institutional Review Board (IRB) approvals (or an equivalent approval/review based on the requirements of your country or institution) were obtained?
    \item[] Answer: \answerNA{}
    \item[] Justification: The research does not involve human subjects,
    so IRB or equivalent ethics approval is not applicable.

\item {\bf Declaration of LLM usage}
    \item[] Question: Does the paper describe the usage of LLMs if it is an important, original, or non-standard component of the core methods in this research? Note that if the LLM is used only for writing, editing, or formatting purposes and does \emph{not} impact the core methodology, scientific rigor, or originality of the research, declaration is not required.
    \item[] Answer: \answerNA{}
    \item[] Justification: LLMs are not used as an original or non-standard component of the research methodology. BitNet~b1.58~2B4T is the
    inference workload evaluated by the proposed accelerator rather than a tool used to develop the research method.
\end{itemize}
\end{document}